# Technology Report : Smartphone-Based Pedestrian Dead Reckoning Integrated with Data-Fusion-Adopted Visible Light Positioning

Shangsheng Wen, Ziyang Ge, Danlan Yuan, Yingcong Chen, Xuecong Fang

*Abstract*—Pedestrian dead-reckoning (PDR) is a potential indoor localization technology that obtains location estimation with the inertial measurement unit (IMU). However, one of its most significant drawbacks is the accumulation of its measurement error. This paper proposes a visible light positioning (VLP)-integrated PDR system, which could achieve real-time and accurate indoor positioning using IMU and the camera sensor of our smartphone. A multi-frame fusion method is proposed in the encoding and decoding process of the system, reaching 98.5% decoding accuracy with a 20-bit-long ID at the height of 2.1 m, which allows the variation in the shutter speeds of cameras and heights of the LED. Meanwhile, absolute locations and step length could be calibrated with the help of a single light-emitting diode (LED), promising average accuracy within 0.5 meters in a 108-meter walk.

## I. INTRODUCTION

NOWADAYS, location-based service (LBS) serve an increasingly crucial role in our everyday lives due to the rapid development of the Internet of Things (IoT) in the era of 5G. Global Positioning System (GPS) could provide accurate locations outdoors. But alternative techniques are required in indoor scenarios, considering the drop in GPS accuracy caused by the signal attenuation and multipath effect [1]. Different approaches have been intensively studied, most of which are radio-based, including Bluetooth, Wi-Fi, Radio Frequency Identification (RFID) etc. [2-4]. The methods above share some common disadvantages, such as electromagnetic interference and extra hardware requirements. Recently, there are two potential methods intensively studied in the area of indoor positioning, visible light positioning (VLP) and pedestrian dead-reckoning (PDR).

There are mainly two types of receivers applied in the area of VLP, including one based on the photodiode (PD) and the other based on complementary metal oxide semiconductor (CMOS). The PD-based positioning methods include the received signal strength (RSS), the angle of arrival (AOA), the time of arrival (TOA), the time difference of arrival (TDOA) etc. [5]. However, most of these methods are environmentally susceptible. More importantly, considering the ubiquity of CMOS-equipped mobile phones and their robustness to the environment, recent trends of VLP favor the latter methods. In the CMOS-based VLP system, data are transmitted through on-off-keying-modulated (OOK-modulated) LED, which flickers at an imperceptibly high frequency. At the receiver end, due to the rolling shutter effect, black and white stripes are formed in the frame captured with controlled exposure, representing '0' and '1' respectively. With LED and CMOS, certain positions are available with a database storing the ID information of each LED. Meanwhile, the shutter speed of cameras and the distance between cameras and LED limit information in a frame, thus needing better demodulating or decoding methods. Furthermore, owing to the limited Field-of-View (FOV) of the camera sensor, the method of VLP depends on the distribution of modulated LED to an extent, making it discontinuous when lacking such light sources. Therefore, another approach is needed to remedy this deficiency.

Dead reckoning has been widely used in positioning. It is a concise method that calculates the relative position using inertial sensors. With the miniaturization of inertial measurement unit (IMU), it could be embedded in different hardware and become available in pedestrian navigation, commonly known as PDR. The two main systems used in PDR are the Inertial Navigation System (INS) and Step-and-Heading System (SHS) [6]. For daily use of pedestrian navigation, the two-dimensional (2D) SHS is sufficient and suitable for smartphone-based applications, because of which it is adopted in our article. The SHS updates the position with the vector {step length, heading direction}. Therefore, the key issues in SHS include step detection, step length estimation, and heading direction estimation. Multiple inertial sensors that include gyroscopes, accelerometers, and magnetometers are used for data collection. And the data is processed for the aforementioned estimation. The method PDR could be completed solely with IMU-embedded mobile phones. Nevertheless, if only with PDR, no global information could be obtained. Another common problem would be the cumulative error, since data collected from the IMU could be inaccurate or even incorrect

This work is supported by the following: The Guangdong Science and Technology Project under Grant 2017B010114001. *(Corresponding author: Shangsheng Wen,* e-mail: shshwen@scut.edu.cn*).*

Shangsheng Wen, Ziyang Ge, Xuecong Fang, Yingcong Chen are with the School of Materials Science and Engineering, South China University of Technology, Guangzhou, China.

Danlan Yuan is with the School of Electronics and Information Engineering, South China University of Technology, Guangzhou, China.

Weipeng Guan is with the School of Automation Science and Engineering, South China University of Technology, Guangzhou, China (e-mail: augwpscut@mail.scut.edu.cn).

sometimes. Faced with such problems, multiple approaches have been put forward. In [7], a method that fused PDR and Bluetooth was raised, allowing localization accuracy of 1 m. Similar combinations could also be found in [8-10], all providing synergistic effects on pedestrian navigation. However, numerous extra hardware devices are usually needed in these methods, leading to extra cost and inconvenience to some extent. Meanwhile, along with LEDs prevalently installed for illumination, CMOS-based VLP could be easily achieved with a modulated LED as the transmitter and a camera sensor as the receiver. This provides great convenience for the simultaneous implementation of the two methods, and we are motivated to combine the two popular localization methods.

The method of VLP would provide global information and periodical calibration for PDR while the application of the inertial sensors perfectly solves the discontinuity problem caused by the limitations of the FOV of camera sensors. Furthermore, we notice that the geometric information captured from the camera sensor could be exploited to improve the step length calibration. To enable longer ID and longer decoding distance in VLP, we creatively develop a method of adaptive multi-frame data fusion.

In this paper, VLP and PDR are tightly coupled to form an integrated scheme, which could be easily realized with prevalently installed LEDs and common smartphones. The two well-developed methods are perfectly complementary in the area of pedestrian indoor navigation. Additionally, the hybrid method of proximity [11] and dead-reckoning is applied in our system, given that centimeter-level precision is usually not required in pedestrian navigation.

We highlight our contributions as follows:

1) The idea of data fusion is adopted in the encoding and decoding processes of the system, allowing a longer ID sequence for each LED and a higher tolerance for the variation in parameters, such as rolling shutter speed of the image sensor and distance between the sensor and LED.

2) The method VLP provides calibrated locations, compensating for the unknown initial state and accumulated error of dead-reckoning.

3) Regarded as one of the key parameters in PDR, step length would be calibrated with the help of the LED in the FOV of our camera sensor, and calibration could be timely implemented with a single LED in a certain area.

The remainder of this article is organized as follows. Section II introduces the related work. Section III details the principle based on VLP and PDR, as well as the overall hybrid system. Section IV describes the set and results of the experiment. Finally, in section V, the conclusion is drawn.

## II. Related Work

### A. Decoding Methods in the Terms of CMOS-based VLP

Apart from positioning accuracy and latency, long distance decoding and higher tolerance in the variation of rolling shutter speed are also key issues in VLP positioning. Usually, in CMOS-based VLP, a bit of data contains several pixel rows. The rolling shutter speed of the camera sensor and the flickering frequency or height of the LED decide the number of pixel rows per bit. If a better demodulating scheme is found, a lower error rate would be realized with fewer pixel rows per bit, thereby enabling greater height of the LED and various shutter speeds. Traditional machine learning demodulating methods have been discussed and evaluated in [11-13]. However, this limit would be reached and could not uproot the problem with only the improvement in demodulating methods. Instead, approaches to increasing the length of the sequence would be needed, and thus the RGB-LED-ID method [15-16] was applied to enable multichannel decoding that tripled the bit per frame. Likewise, the idea of fusing data from different frames has been previously adopted in image processing and communication [17]. We adopt and improve this idea in our system, which allows data to be split into parts and thus robust in the variation of shutter speed and distance.

Another problem faced in the VLP system is the distribution of the modulated luminaries. Given that the LED could be sparsely distributed, the geometric feature of a single LED is exploited to achieve three-dimensional positioning. Aiming to compensate for the reduced number of LED luminaries, Zhang et al. [18] used an extra point marker on the lamp to perform circle projections and avoided using angular sensors. In [19-20], information captured with IMU was combined to exploit projective geometry. Despite their high accuracy, they were still limited to a relatively small space because of the limited FOV of the camera and had slightly different practical applications. To achieve positioning in a vaster indoor scenario, VLP fused with other methods which have been intensively studied in the area of robotic localization [21-22, 32-34], proving the feasibility and robustness of sensor fusion.

### B. Methods in Terms of SHS

IMU units embedded in smartphones are of relatively poor performance. What's more, it is not placed in the mass of the body most of the time. Therefore, most conventional estimating algorithms that are suitable for mounted devices have a poor performance on smartphones. As mobile phones prevailed, methods were developed to cope with smartphone-based scenarios.

Among all step detection algorithms, peak detection has the best performance and is applied in most PDR systems [6]. In the case of IMU embedded in mobile phones, it is hard to avoid occasional errors as people walk in different manners. A motion mode classifier was adopted in [23] to detect step events correctly.

As for step length estimation, the Weinberg model [24] is commonly used, with the maximum and minimum of the perpendicular acceleration $a_{zmax}$ and $a_{zmin}$ derived to estimate the length,

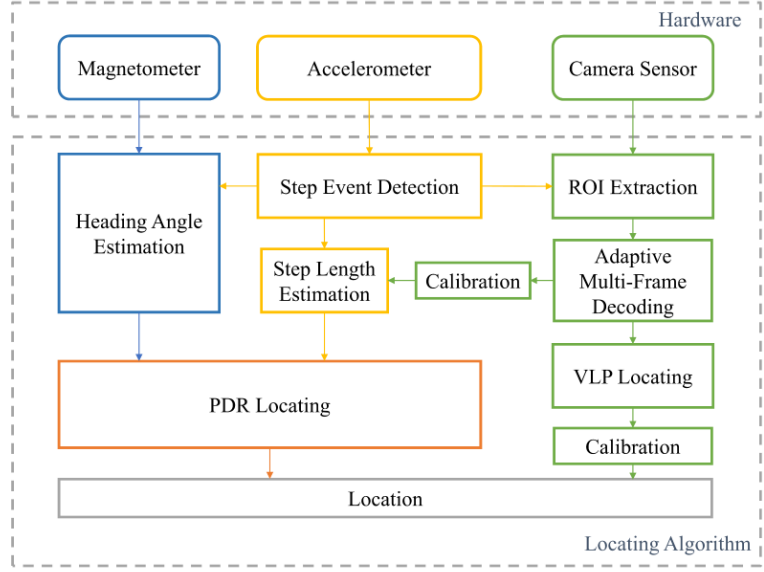

**Fig. 1.** The diagram of our VLP-integrated PDR system

$$l_1 = k_1 \cdot \sqrt[4]{a_{zmax}^2 - a_{zmin}^2}, \quad (1)$$

where $k_1$ is a constant used for unit conversion.

A method to improve the Weinberg model was proposed in [25], where an adaptive value related to velocity instead of a constant replaced the unit conversion. In [26], step frequency $f$ and stature $s$ were used to estimate step length,

$$l_2 = k_2 \cdot f \cdot s, \quad (2)$$

where $k_2$ is a constant used for unit conversion.

Visible light was applied to assist the inertial navigation in multiple articles before. In [27], trilateral RSS was used in VLP and fused with PDR with an extended filter, requiring the modulated LEDs to be densely distributed. In [28], Xu *et al.* proposed the IDyLL system, which fused data from IMU and photodiode using a particle filter, while the distance between lamps was used for step length calibration. Nevertheless, the method for calibrating the step length between LEDs would limit the distribution of modulated LEDs, otherwise, calibration would be inaccurate if the pedestrian is not walking straight between the luminaries. In [29], CMOS-based VLP was applied to calibrate the PDR-acquired location, while step size and step number between LEDs were used for verification. However, step length was estimated only with the Weinberg model, which was not sensitive to the change in speed and individual difference. With the aim of accurately locating pedestrians in a vaster indoor area, an idea of calibrating step length with a single LED is proposed in our system.

III. PRINCIPLES OF THE PROPOSED METHOD

In this section, we propose a detailed method of PDR tightly coupled with VLP and an encoding method which is suitable for this system.

*A. System Overview*

Fig. 1 is the diagram of our integrated system with VLP and PDR which are tightly coupled and highly complementary. An adaptive multi-frame fusion method was proposed in VLP and a step length calibration method was raised. We assumed that a certain number of modulated LED luminaries would be distributed in a certain area. And the pedestrian would hold the phone horizontally without tilting with the direction pointing to the heading direction while walking in the area. 2D navigation could be realized by our proposed system.

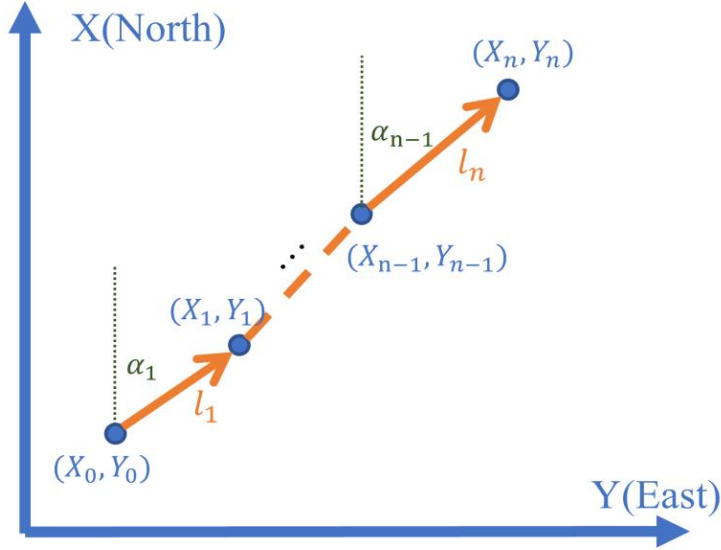

**Fig. 2.** The illustration of PDR localization

*1) VLP Basis*

For the VLP part, each frame of image captured with fixed exposure time is preprocessed for ROI extraction. Firstly, the captured frame is binarized and performed close operation. In the captured frame, black stripes are eliminated to form a white solid circle that is the region of interest (ROI) in the VLP system. Then we obtain the center and the size of the region. In the decoding process, the counterpart of the ROI is cropped out from the unprocessed image and decoded by the efficient decoding scheme [14]. The data of each frame is collected and gathered to acquire the ID of LED by using the adaptive multi-frame data fusion method. We store the size and position information of the lamps corresponding to the IDs in the database in advance.

*2) PDR Basis*

For the PDR part, we adopt the SHS system. As is stated in [30], step detection is the most accurate with accelerometer, and thus step event is detected with acceleration using the method of peak detection in our system. Every time a step cycle is detected, the PDR system outputs the updated 2D coordinate with step length estimated and heading direction obtained by accelerometers and magnetometers. The updated coordinate is calculated as

$$\begin{bmatrix} X_n \\ Y_n \end{bmatrix} = \begin{bmatrix} X_{n-1} \\ Y_{n-1} \end{bmatrix} + l_n \begin{bmatrix} sin\alpha_n \\ cos\alpha_n \end{bmatrix} \quad (3)$$

where $\alpha$ is the angle between the estimated direction and the north, measured clockwise around the pedestrian's horizon and the heading direction, $l_n$ is the estimated step length, $(X_n, Y_n)$ is the current position, $(X_{n-1}, Y_{n-1})$ is the position of last step event. The process is also illustrated in Fig. 2.

*3) Integrated Algorithms*

A default position would be given in the first place, and step length would be calculated with acceleration. Then absolute position along with global information would be acquired, and step length would be calibrated once the pedestrian walks past the circular LED luminary. If the LED is out of the FOV of the camera sensor, the system would work in the mode of PDR, using the step length calibrated with VLP.

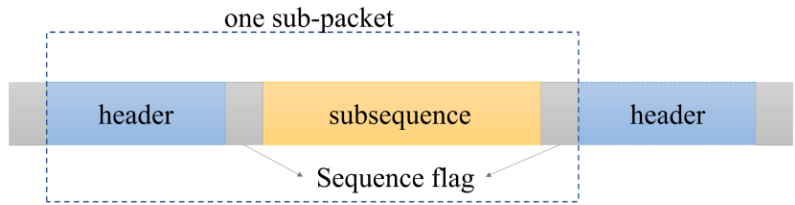

**Fig. 3.** The structure of sub-packet we adopted in our adaptive multi-frame decoding method

*B. Adaptive Multi-Frame Data Fusion*

Here, we adopt the approach of data fusion to obtain the ID of LEDs of greater heights and with camera sensors with various shutter speed. In traditional methods, a data packet usually consists of the payload and the header and is sent repeatedly. At the receiver end, after locating the header, the data between the headers would be extracted and decoded.

In our proposed method, sequence flags are added at two ends of the payload. The flags are used to mark the sequence of the payload, allowing data fusion within and between frames. The structure of each sub-packet is illustrated in Fig. 3. At the receiver end, each captured frame is firstly demodulated into binary sequence. Then, aiming to decode, the headers are searched, and the sequence flags are recognized. We calculate the number of "0" and "1" at each position and do majority voting every few frames, where the number of frames required would vary adaptively with the change of data rate. After that, all data sequences are decoded and put together according to the sequence flags to get the full ID sequence, as illustrated in Fig. 4. The aforementioned adaptive frame interval is decided with the following method.

We count the number of valid bits excluding the header and the sequence flag per decoded frame and denote it as $b_n$. The counted bits of the unselected frames are accumulated and denoted as $b_{cnt}$,

$$b_{cnt} = \sum_{n=1}^{k} b_n, \quad (4)$$

where $k$ is the number of the unselected frames while $n$ is the index of frame in each round.

We set a reference boundary denoted as $b_{ref}$ according to the performance of the camera sensor and the scenario required. In addition, we define the full length of our predetermined ID as $b_{full}$. When the average number of bits per frame reaches the length of ID, the data is enough for immediate decoding. Otherwise, we would wait until enough bits are accumulated to ensure every data bit is included, thereby setting the majority voting threshold as

$$b_{th} = \min(b_{ref}, k * b_{full}) \quad (5)$$

Once $b_{cnt}$ becomes greater than or equal to $b_{th}$, majority voting would be performed on the accumulated data, and the ID information would come out immediately. The value k at the moment would be defined as the frame interval between the two data gatherings of this round. Then the variables are reset and the algorithm moves on to another round of accumulation. With the adaptive method of frame interval selection, short-distance decoding could be accelerated and long-distance decoding could also be guaranteed at a relatively low speed. Apart from the tolerance of greater distance and camera sensor performance, multi-frame data fusion also has better stability considering the repetition of the number at each position compared to single frame methods.

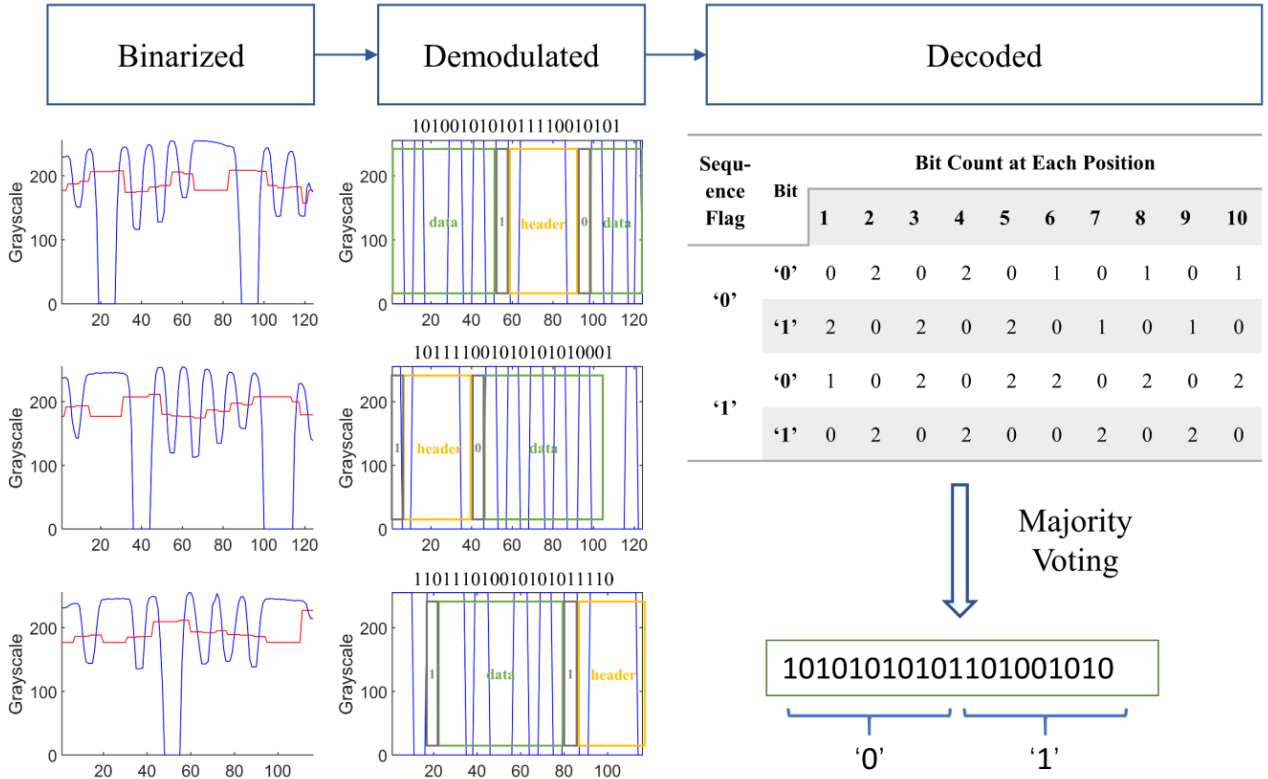

Fig. 4. Illustration of the decoding process using majority voting (The blue curve on the leftmost row represents the grayscale value with background noise removed, and the red curve represents our calculated threshold)

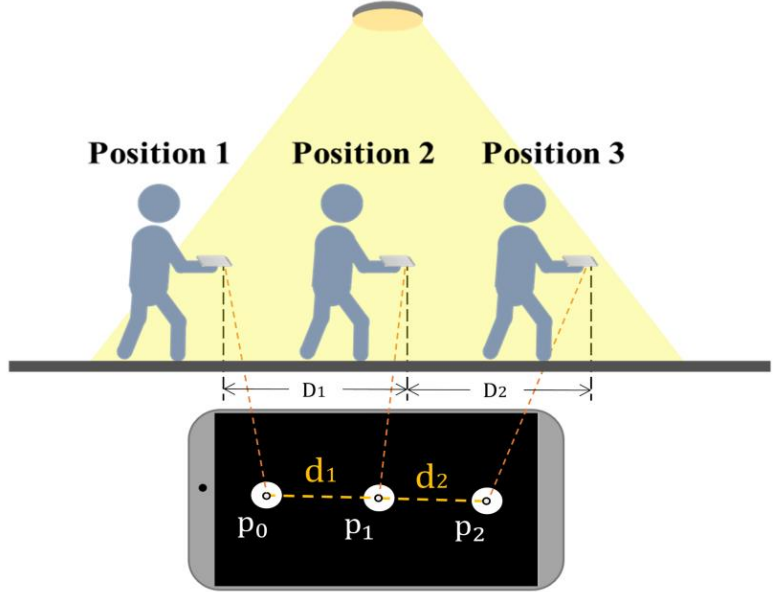

**Fig. 5.** The illustration of step length calibration

*C. Step Length Calibration*

As is known to all, step length is a key parameter in the process of dead reckoning. Meanwhile, such parameter varies with individual stature, walking habits and even walking speed. However, the sensors in IMU are not always accurate, making it unreliable to acquire definite data with IMU only. On the contrary, such data could be accurately obtained via the camera sensor. As the pedestrian moves forward, the LED would relatively move backward for a certain distance. With VLP assisted, step length could be calibrated according to such distance.

When the pedestrian is under a certain LED, as step event detected, the center of the ROI in that specific frame is recorded as $p_n$, and the corresponding coordinate would be $(x_n, y_n)$. The Euclidean Distance between the adjacently recorded centers is calculated, defined as $d_n$.

$$d_n = \sqrt{(x_n - x_{n-1})^2 + (y_n - y_{n-1})^2} \tag{6}$$

Knowing the actual radius $R$ of the LED and the radius r of ROI in the captured frame, we could estimate the moving distance of the camera sensor which fully represents the step length of the pedestrian, as shown in Fig. 5.

$$D_n = \frac{r}{R} d_n \tag{7}$$

However, there is another situation to consider. Given that smartphones are usually held a distance away from the body of the pedestrian, the ROI in the frame could drift a considerable distance when people make a turn, even if the pedestrian is at the same point, which would severely affect the calibration. Thus, we define a new variable $\Delta\alpha$ as the turning angle, and set a threshold as $T_\alpha$.

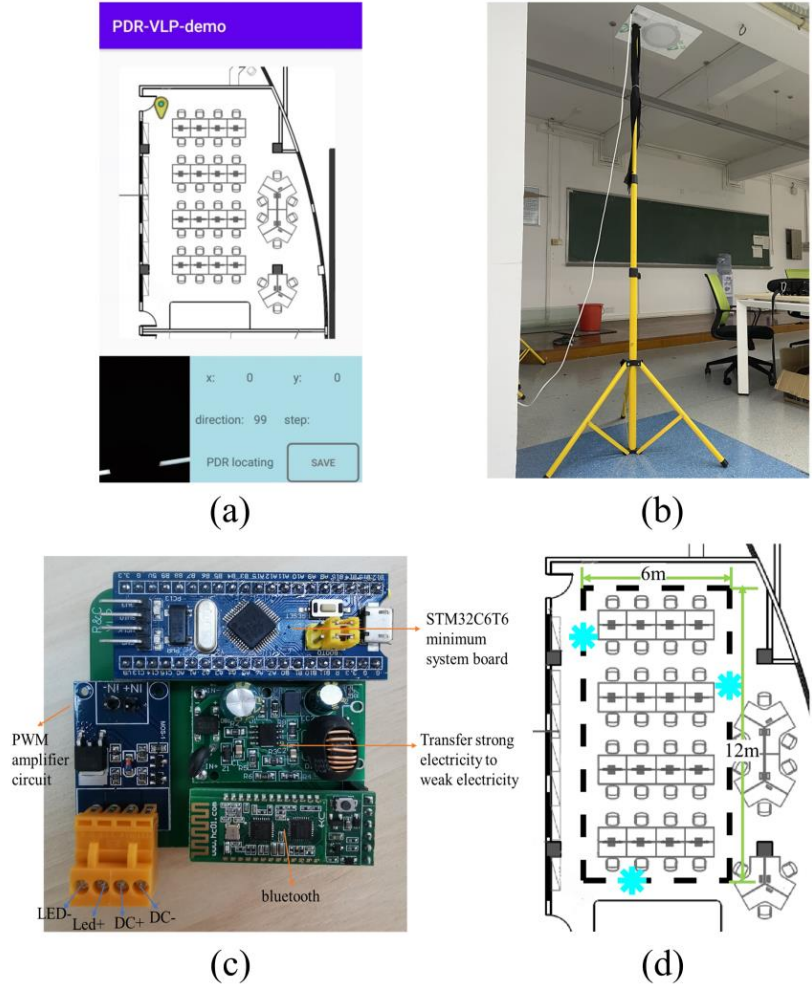

**Fig. 6.** Experimental setup (a) Android application, (b) the modulated LED, (c) tailor-made PCB and (d) planned route for the experiments

$$\Delta\alpha = \alpha_n - \alpha_{n-1} \tag{8}$$

If the turning angle $\Delta\alpha$ in one step reaches $T_\alpha$, the step length data of that frame would be discarded with the frame index recorded in set $K_d$. Therefore, the corrected step length under the certain LED could be calculated as:

$$D_{av} = \frac{1}{k} \sum_{n=1, n \notin K_d}^{k} D_n \tag{9}$$

If we consider the variation between steps, the Weinberg model is adopted with the value $k_1$ updating under each LED. And our value $k_1$ is calculated as:

$$k_1 = \frac{D_{av}}{\sqrt[4]{a_{zmax}^2 - a_{zmin}^2}} \tag{10}$$

*D. Proximity-Dead-Reckoning Hybrid Positioning*

There are several common algorithms in the area of localization and in [5], they are classified as fingerprinting, triangulation, proximity, vision analysis, and dead reckoning. With proximity-based positioning, the coarse location would be acquired from the signal of a single LED. The location will remain the same if the LED is in the FOV of the camera sensor. A relatively low accuracy would be obtained with such a method. However, it is still of great practicality in quite a few scenarios. Fused with dead-reckoning algorithm, location information would be obtained from IMU in the area where the proximity method is unavailable, ensuring the continuity of the positioning system. Combining two localization methods, this method would be quite enough to satisfy the need for pedestrian navigation even if the modulated LEDs are sparsely distributed.

IV. EXPERIMENT

We conducted several experiments to verify the performance of the proposed application of multi-frame data fusion and the proposed system of PDR tightly coupled with VLP.

*A. Experimental Setup*

*1) Hardware*

To make luminary modulation more convenient, we designed a tailor-made Printed Circuit Board (PCB), as shown in Fig. 6. The Bluetooth on the PCB can communicate with an Android-based application, transmitting the information to the Microcontroller Unit (MCU), namely, STM32C6T6, which controls the on-off state of the LED through Pulse Width Modulation (PWM).

The mobile device used was a Huawei Mate 10 Pro, while the front camera was accessed throughout the process. The exposure of the camera sensor is fixed for demodulating and the frame rate is set at a maximum of 15fps. Meanwhile, the sampling rate of the accelerometer was set as 250sps. In our experiments, all data capturing and processing are done on the mobile device; thus, it would not perform as fast as data calculated on the computer.

*2) Software*

We built up an Android application to perform our experiment and the layout is shown in Fig. 6. For the main part of the app, a stretchable map of the laboratory is shown and the green locating icon on the map indicates the current location. The captured frame is shown at the bottom left, while other details are placed at the bottom right. Additionally, the save button is used for data

TABLE I
ADAPTIVE MULTI-FRAME PERFORMANCE

| $b_{ref}$ | height(m) | accuracy(%) | latency(ms) | $b_{ref}$ | height(m) | accuracy(%) | latency(ms) |
|---|---|---|---|---|---|---|---|
|  | 2.2 | 47.0% | 490 |  | 2.2 | 86.4% | 525 |
|  | 2.1 | 94.0% | 452 |  | 2.1 | 94.5% | 491 |
|  | 2 | 97.2% | 410 |  | 2 | 99.3% | 459 |
|  | 1.9 | 98.0% | 373 |  | 1.9 | 99.7% | 421 |
|  | 1.8 | 99.5% | 350 |  | 1.8 | 99.8% | 387 |
|  | 1.7 | 100.0% | 275 |  | 1.7 | 99.8% | 370 |
| 50 | 1.6 | 99.8% | 268 | 70 | 1.6 | 99.7% | 341 |
|  | 1.5 | 100.0% | 263 |  | 1.5 | 99.9% | 330 |
|  | 1.4 | 99.7% | 232 |  | 1.4 | 99.7% | 319 |
|  | 1.3 | 99.9% | 209 |  | 1.3 | 99.7% | 292 |
|  | 1.2 | 100.0% | 187 |  | 1.2 | 99.8% | 289 |
|  | 1.1 | 99.9% | 100 |  | 1.1 | 100.0% | 118 |
|  | 1 | 99.9% | 73 |  | 1 | 100.0% | 74 |
|  | 2.2 | 63.0% | 544 |  | 2.2 | 89.0% | 611 |
|  | 2.1 | 94.2% | 540 |  | 2.1 | 98.5% | 567 |
|  | 2 | 98.0% | 396 |  | 2 | 99.5% | 526 |
|  | 1.9 | 99.0% | 370 |  | 1.9 | 99.5% | 507 |
|  | 1.8 | 100.0% | 336 |  | 1.8 | 99.8% | 451 |
|  | 1.7 | 99.6% | 317 |  | 1.7 | 100.0% | 412 |
| 60 | 1.6 | 99.8% | 297 | 80 | 1.6 | 99.8% | 390 |
|  | 1.5 | 99.8% | 294 |  | 1.5 | 99.7% | 383 |
|  | 1.4 | 100.0% | 289 |  | 1.4 | 99.9% | 366 |
|  | 1.3 | 99.9% | 286 |  | 1.3 | 100.0% | 333 |
|  | 1.2 | 99.8% | 280 |  | 1.2 | 99.8% | 321 |
|  | 1.1 | 100.0% | 103 |  | 1.1 | 99.8% | 135 |
|  | 1 | 100.0% | 73 |  | 1 | 99.9% | 72 |

saving, exporting key information to Excel for later evaluation.

TABLE II
SINGLE FRAME DECODING PERFORMANCE

| height(m) | accuracy(%) | latency(ms) |
|---|---|---|
| 1.6 | 54.69 | 79ms |
| 1.5 | 97.7 | 72ms |
| 1.4 | 99.05 | 70ms |
| 1.3 | 99.8 | 75ms |
| 1.2 | 99.95 | 80ms |
| 1.1 | 99.9 | 67ms |

*B. Multi-Frame Fusion Performance*

For multi-frame fusion, the experimental equipment is set up as Fig. 6. Multi-frame decoding and single-frame decoding are, respectively, performed to draw a contrast. In the two sets of experiments, the flickering frequency is set as 15000Hz, meanwhile, the coding manners are similar, to the same header '011110' and the same payload that contains 20 bits of data. The 20 bits would be separated into two groups of 10 bits in multi-frame encoding.

For multi-frame encoding, the sequence flags are set a one bit long, which are '0' and '1' respectively. We measured the accuracy and latency at different heights and with different $b_{ref}$ values. Our results are shown in Table I. For single-frame scheme, accuracy and decoding rate are also measured at different heights, which are recorded in Table II.

With our proposed method, a decoding accuracy of 98.5% could be reached at a decoding distance of 2.1 m and an accuracy of 89.0% at 2.2 m when the $b_{ref}$ is set as 80. Meanwhile, with the same data payload setting, similar accuracy of 97.7% could only be reached at 1.5 m and deteriorates quickly as the distance lengthens using the single-frame decoding method. Additionally, with the adaptive method, the multi-frame decoding has a latency performance getting increasingly close to the conventional single-frame performance as the decoding distance shortens. The latencies are both in the vicinity of 70ms when the distance is 1 m.

However, as can be seen from the data, the multi-frame fusion method has a significant increase in latency compared to the single frame method at around 1.5 m. Extra headers and sequence flags needed for multi-frame fusion with the same payload cause that. Furthermore, the frame rate of the camera sensor and the calculating performance of the device limit the decoding rate severely. Combined with our proximity method used in pedestrian scenarios, such latency is acceptable.

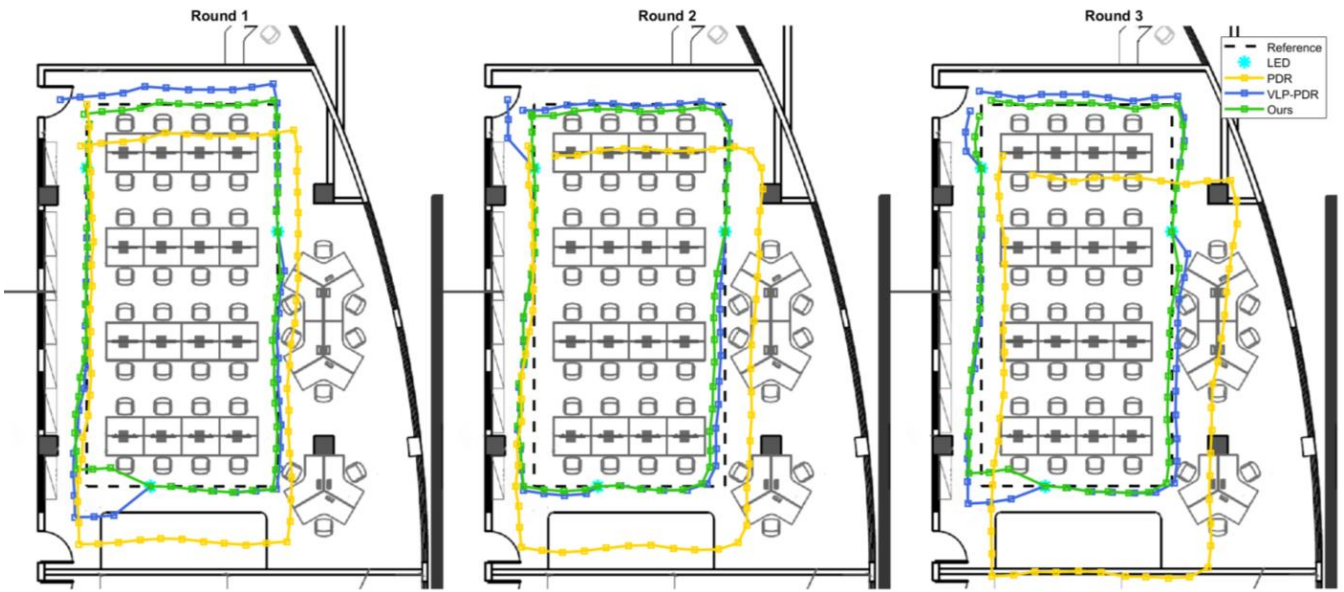

**Fig. 7.** The trajectories of the three positioning methods of each round at relatively low speed (0.9~1.2 m/s)

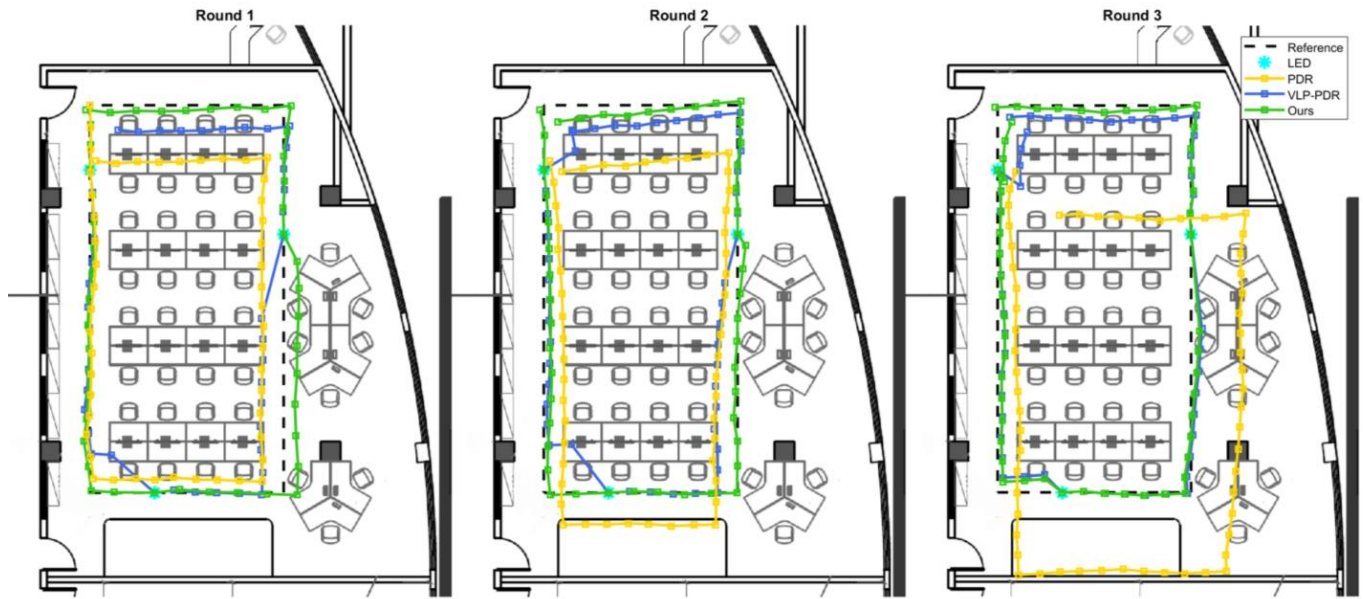

**Fig. 8.** The trajectories of the three positioning methods of each round at relatively high speed (1.4~1.7 m/s)

*C. Localization Performance*

*1) Experiment*

For test operations, we equipped our laboratory with three LED luminaries. The luminaries were modulated with the method of multi-frame fusion, and the IDs were correspondingly stored in the database in advance. The modulated luminaries and the mobile application of the system are set in the same way as in part A of this section and distributed as in Fig. 6. Our route is set as a rectangle that is 12 m $\times$ 6 m with the three modulated luminaries equally distributed, which means we would walk past the LED every 12 meters. Three rounds around the rectangular path would be completed in each experiment, so our route would be 108 m.

To test the performance of our proposed hybrid system, we performed our experiment at various walking speeds with three positioning methods simultaneously. All methods are based on the Weinberg model mentioned in equation (1). The initial constant in the model used for comparison is referred to [30] and then slightly adjusted according to the actual step length we tested while walking at normal speed around 1.2 m/s. The PDR algorithm has no calibration, whereas the VLP-PDR method has the absolute position calibrated below each modulated LED. In our proposed method, not only the absolute position but also the $k_1$ values in the Weinberg model are calibrated. The routes and the test results are shown in Fig. 7 and Fig. 8, which are performed at a relatively slower speed (0.9-1.2 m/s) or, respectively, higher speed (1.4-1.7 m/s).

Additionally, considering that the actual length of every single step is unknown, we evaluated the system by calculating the error every round at the starting point, comparing to the position of the reference starting point. We repeated the experiment 30 times at various walking speeds and took the average.

*2) Results and Discussion*

Comparing the two figures, it's clear that the shape of the VLP-PDR trajectory (blue) changes with the walking speed, with convex at corners at lower speed and concave at higher speeds in contrast to the reference route. As for the PDR trajectory (yellow), the triangle would be smaller or larger than the reference. The changes in step length at various walking speeds cause such errors. Step length usually lengthens as the speed goes up, and extra distance would be counted if walking at high speed, as shown in Fig. 8. Furthermore, for situations where only PDR is applied, the misjudging of step event would lead to a shift of the whole trajectory, as could be obviously shown in the third round of both figures.

As can be observed from the trajectories, the calibration of VLP dramatically improves the performance of the positioning system. It eliminates the accumulated positioning error every certain distance and thereby constantly keeps the accuracy within 3 m even without the step length calibration implemented. With the calibration of $k_1$ value in step length estimation performed, a slight increase in accuracy could be seen due to the improvement in sensitivity of variation in walking speed.

The positioning error of each round is shown in Table III. The error increases when using PDR only, whereas the accuracy of the latter two methods of each round maintains at a similar value with the calibration of the absolute position performed. Additionally, our proposed method would be slightly better with the step length calibrated.

TABLE III
AVERAGE LOCALIZATION ERROR AT THE STARTING POINT

| Method | PDR (m) | VLP-PDR (m) | Ours (m) |
| --- | --- | --- | --- |
| Round1 | 1.421 | 0.804 | 0.501 |
| Round2 | 2.416 | 0.754 | 0.509 |
| Round3 | 3.742 | 0.766 | 0.475 |
| Average | 2.526 | 0.775 | 0.495 |

All the data above indicate that our proposed hybrid system would guarantee user experience with a better performance in accuracy and availability in an LED-sparsely-distributed scenario. Furthermore, with VLP implemented, global information would be obtained, which properly complements the PDR method that could only acquire a relative position.

## V. CONCLUSION

In this paper, we propose a VLP-integrated PDR system, which could provide real-time local positioning with a common cell phone in an area where the LED luminaries are sparsely distributed. The applied multi-frame fusion scheme enables higher tolerance in the height of LED and variation of shutter speed, allowing 98.5% decoding accuracy with 20-bit-long ID at a height of 2.1 m. As for our hybrid system, we could adapt to the change in walking speed and reach an average accuracy within 0.5 meter each round. The results show that our method could be practical in pedestrian navigation due to the following reasons. Firstly, our method requires only a common smartphone and LED, which would be easy to implement and cost-saving. Secondly, longer ID is allowed in multi-frame fusion and LEDs could be sparsely distributed, promising locating in vast indoor scenarios. Lastly, it is reliable and practical for its robustness and submeter accuracy in positioning. For a more accurate VLP-PDR-integrated positioning system, we expect to combine orientation calibration or motion classification with this method, which will our future explorations.

## ACKNOWLEDGMENT

This work is supported by the following: The Guangdong Science and Technology Project under Grant 2017B010114001.

## OUR WORK IN VLP

We start our research in the field of VLC and VLP since 2014, we have published many research articles in prestigious international journals and conferences. The representative works from our group can be seen in Ref. [35-37], which is about VLP for high accurate robot indoor localization using RSE-based Camera. Besides, we also conduct many works for the PD-based VLP using optimization algorithm (can be seen in Ref. [38-50]). Apart from the VLP, we also deeply explore the OCC method for IoT application or reliable wireless communications [51-64]. A presentation about our work in the field of high accuracy robot localization using VLP can also be seen in[1]. For the readers that get insights from our research works, please cite our works.

---

[1] https://b23.tv/FWQWPsg